# Concepts as Elementary Constituents of Human Consciousness.

Boris Rusakov
*(rusakov@xpertnetinc.com)*

**Abstract.**

It is asserted that consciousness functionally is a vision. Biologically it is a sensation of work of a brain that converts external and internal signals into visual output. However, as we well know, human consciousness includes meaning (interpretation), and therefore functionally possesses an additional level of complexity. It consists of concepts which are its minimal components (elementary constituents). A concept is an abstract (i.e. non-existent in nature) common property of different observables, as perceived by humans. It is an irreducible entity since it cannot be divided into any other functional components. At the biological and physical level each concept is a unique sensation encoded in a nervous system as a sensory and visual image. Concepts evolve throughout one's lifetime. We continuously create and acquire new concepts, develop and expand the existing ones, and eliminate and abandon some others. 'I' ('self') is one of the concepts. To acquire a concept one has to learn (to teach one's nervous system) feeling or sensing it, and to connect it to existing concepts, if any. We make a conjecture that both the creation and the acquisition of a concept is a phase transition. The ability to generate and acquire concepts is unique to humans and is the only principal difference between human and animal. We offer scenarios for how this ability could have been acquired by our predecessors, thus making them humans. It is suggested that the animal's brain, being a processor of visual signals that converts them into internal images, has developed this ability as a result of "guessing" or imagining details that are missing in the actual visible picture.

## Introduction: Concepts.

Looking at an oak, a birch, a pine or a willow, animals unlike humans do not try to understand what is common between them. They do not decide to call them all 'trees' and do not look for what could be other trees. They do not analyze, systemize, extract common properties, or look for reasons and explanations. Humans do. Why there is such a difference? When did it start? Below, I hope I will answer these and some other questions.

Extracting common property is what we call conceptualization. The 'tree' in this example is a concept. It is an abstraction that does not exist in the real world. It is an abstract figure of an imaginary generalized plant having distinguished stem, branches, sub-branches, and possibly leaves (though leaves are not always necessary). A concept is an abstract common property of different observables (values), their common classifier, common description or feature. The crucial property of a concept is its independence from the specific object from which it was singled out, its portability to other objects. We even apply it to completely unrelated subjects. For example word "tree" as used in the phrases "tree of life", "family tree", "tree-like diagram" has nothing to do with the plant it has originated from. In mathematics we would say that the concept is a vector, i.e. a collection of values, or components. In the example above, oak, birch, pine, and willow, are specific values of the concept of a 'tree'. Though oak, pine and the rest of them are the concepts themselves, since they are generalizations of their own sub-categories, such as red oak, white oak, poison oak, etc.

Everything our consciousness operates with is a concept. Any word or phrase denotes a concept. Any action, verb, logical operator, or mathematical construct is a concept as well. To sit, to eat, to think, to imagine are concepts. When we say "sit" what comes to the mind? It could be several images or clips simultaneously. It could be a person taking a sit on a chair, or a person doing squat, or a dog performing the corresponding command. In any case it is an image or a clip of someone bending their limbs and lowering their posture, most likely to lower their rear to a hard surface. In itself, it is an abstraction detached from any particular reality, while taking values in multiple particular realities.

Our definition of concept does not contradict to its commonly adopted definitions as "something conceived in the mind: thought, notion" (Merriam-Webster) or "an abstract idea, a general notion" (Cambridge Dictionary) and etc. We just realized that it applies to everything our consciousness operates with and consists of, including a concept of 'consciousness' itself.

In other words, concepts **form** human consciousness. An attempt to divide a concept into more fundamental constituencies is fruitless. The concept is irreducible entity. It can only be described in terms of other concepts, if any, which in itself is a proof of its irreducibility. It is just like "elementary particles" (another concept) in physics. They cannot consist of anything, and can only react with each other and produce other elementary particles. That is why we call them elementary.

Any concept at the same time could be a value of another concept as we just saw. Conceptualization is an abstract thinking since concepts do not exist in nature. They are only in the human mind, and are acquired in the process of learning. Moreover, even in the human mind, they do not appear at birth.

We are born with ability to learn concepts but we do not have a single concept at birth. It takes time for a baby to acquire such concepts as 'mom' or 'dad'. A concept of 'I' appears significantly later, and develops and expands throughout one's lifetime.

Biologically, each concept is a unique sensation, a sensory image of corresponding unique biochemical profile, or state, or excitation, in the individual's nervous system. And this is how it is stored in the nervous system. It is also a visual image.

Concepts change with the time as our experiences and education change. Some of them become more sophisticated. Others fade away or become obsolete because of their irrelevance. Many concepts have disappeared as contradictory to science. Many were dismissed as inadequate. And certainly we acquire new concepts. This is not easy to do. To acquire a new concept we need to train our nervous system sensing it, to relate it to other concepts we previously acquired, to accommodate it into existing system of concepts, i.e. into consciousness.

Another important property of a concept is that it cannot exist without notation. Any new concept at the time of its birth must be denoted. The notation could be a picture, a sign, a sound, a meme, a gesture, or a word or phrase. The notation is what allows us to transfer a concept from one object to another. In our civilization the main instrument of notation is language. The word or phrase denoting a concept is its integral part. I am far from claiming that absence of language automatically indicates absence of concepts. As mentioned above, there are other ways of notation that can be used. However, language is the most versatile and dynamic one, and we are very lucky to have it.

Since every nervous system evolves over time and since there are no two identical nervous systems, it is impossible that the same concept is felt identically by two different people or even by the same person at different moments in time.

The privacy of consciousness is a direct result of this uniqueness. Only the owner knows about his or her feelings and sensations. We cannot read another person's thoughts. We cannot even determine whether we see a color in the same way. Actually, I claim just the opposite – there are no two individuals who see a color in the same way. While it is easy for us to establish that a color-blind person is such, since he does not distinguish one color from another, we cannot find out how exactly another person, who is not color-blind, sees (or better say feels) a specific color. The concept of color has no meaning in physical world. What we associate with a 'green light' is an electromagnetic wave of wavelength of 550 nm. It has no "color". The green color is the sensation, the internal image, produced by our nervous system as a response to this electromagnetic signal. It is clear that there are no two people who see it in the same way, as every nervous system is different.

When something hurts, your doctor doesn't "see" the pain. That is why he asks you to describe it. The only things your doctor can "see" using the most advanced tools is where you have an inflammation or which parameters are abnormal. From the experience with other patients (s)he may be able to assume that you feel a pain, but (s)he cannot be sure that you do. Equally, none of a person's sensations can be directly seen or detected by another person.

**Sensation is a reaction of a body to external and internal signals received by its nervous system.** It is both a state and a never ending process since internal signals are continuously produced by nervous system as a response to its own signals. They are real physical phenomena despite that concepts (being encoded in sensations themselves) are associated with abstractions.

**Side note on Memory.**
I am quite sure that the following happened to almost everyone: you forgot a familiar word or familiar name, and are trying to recall it. In order to recall you are scrolling through suggestions brought up by your mind and dismiss them one by one. You know exactly when it is not it, unless it is it. When the suggestion is close you hesitate for a few seconds, but then decide that it is not it. Indeed, when it is it, you would not hesitate, you would know it immediately and unmistakably. But how do you know that, if you forgot it? Answer: It is because your body remembers the exact sensations in which it was tabulated. When the sensation is not exactly the same, you know it, even if it is very close. Thus, memory of any concept, any word, is a unique sensation corresponding to it. Our nervous system remembers a chemical portrait of the whole concept. We may forget its verbal notation, i.e. the word associated with it, but the chemical portrait is still there and comes up when we recall the notation.

In addition to the privacy of consciousness, another reason why it is so hard to study is our language that we use to deceive ourselves and others in everything related to consciousness. When we say "I",

"I think", "thought", "soul", etc, we think we know what we are talking about, but we don't. How exactly do we think? We don't know. Then how can we claim we do something that we cannot even remotely describe how it's done? These are the concepts created by our consciousness to avoid understanding of what is really going on in it. The first step in gaining this understanding is therefore to expose the deception and to dismantle the perception of consciousness as something fundamentally mysterious.

Despite the deception, the achievements of human intelligence deserve our admiration. Thanks to them we have our cities, roads, books, atomic energy, ships, cars, planes, etc. Actually, it is not despite the deception, it is because of it. By the way, this echoes what Harari described in his book [1]. It is amazing how well it works, that we can successfully use our consciousness without caring at all about what it is and what happens in it. So we will assume that so far it has been a deception for the good. Except that not all achievements of the human mind were peaceful.

Our "I" (Self) is just another concept that we identify with a mysterious owner and controller of our consciousness. But in fact it does not even exist. Imagine a theater without a director and without a script. It's just the actors and the spectators. The actors do whatever they want in order to please the audience and to gain as much applause as possible. As a result, depending on actors' talent and on spectators' taste, there will be some kind of action, sometimes good, sometimes not. It gets even more interesting if actors and spectators are same people. As well, we can add next layer of spectators watching the original spectators and applauding them, and so on. The action may become even more interesting. But in any case, an outside observer would never guess that there is simply no director and script behind all this. This analogy is very close to what happens in our consciousness. Our "I" being this inner producer, even though absent as an entity, is still there as a concept of "would-be" producer. In daily life "I" allows us to bypass complicated details of what is going on in the brain. Try to remove the word "I" from your vocabulary and you will see what a nightmare it saves us from. This is equally applicable to all other concepts related to it, such as "I think". It is clear that each of our actions, each word, logical construct, is a concept.

The presence of an apparatus of concepts (or module of concepts) implies abstract thinking, i.e. ability to separate a property from an object and treat it as a separate entity. A direct consequence of abstract thinking is associative thinking which is a reverse operation, when thinking of one phenomenon leads by one of its concepts to unrelated phenomenon that incidentally has the same common index (a concept).

## Consciousness is a vision and a sensation of the brain work.

The brain and the nervous system in general is a processor of external and internal signals. This is its biological function found in all animals with developed nervous system. The task is to observe, detect changes, and provide the rest of the body with an adequate response to the observed events. While eyes, ears, nose, etc are our sensors, **the brain is the internal display organ** that converts them into internal "images". As mentioned before there is nothing "green" in a green light, it's just an electromagnetic wave of certain wavelength. Similarly, there is no smell or taste in the molecules we eat. There is no "music" in oscillations of air density our ears detect. All of these images, smells, sound, and tastes are the results of work of the brain. All of its output is visual, including thoughts.

Therefore, functionally, **the Consciousness is a vision**.

Any thought is an image, a picture, a clip. Sound and smell are instantly translated into a picture. Pictures are constantly compared with the existing templates. But while animal's brain function is limited to this task, a human's one has "conceptualized" part, which apparently came into existence due to surplus of biomaterial needed for observation. As a result, it has extended its function to guessing, imagining "missing" parts in order to "complete" the picture, i.e. inventing what is not there, speculating, deceiving and lying (on purpose or not), coming up with theories and explanations.

In both human and animal brain, there is a constant flurry of work on the exchange of signals between neurons. The main goal of this work is to display a picture of the "seen" and provide the body with a reaction to it. I use quotation marks here because it doesn't really matter how much of the "picture" came from the outside source. The "complete picture" is anyway generated by the brain itself. For as long as brain is alive this work does not stop, even in a dream or a coma. The only thing that separates the human consciousness from the animal one is abstract thinking.

The electrochemical processes, the excitations of various parts of the brain, are observed and well studied by researchers of the brain and experts in mapping the brain, see for example [2]. But at the same time, the pictures that appear in the brain as a result of the sensation of these signals by the body cannot be seen, since an outside observer does not and cannot know, what sensations this or that signal causes in a particular individual. It depends on the individual's history, experience, education etc. Nor is it possible to directly see the pictures they evoke. Therefore, it is not surprising that researchers who study the work of the brain attribute a certain mystery to consciousness, as if it is something intangible, not directly connected to brain, or even if connected, then in some superficial way.

Cinema is a good illustration of how it works. When we watch a movie in the cinema we know very well it is just a flat screen that reflects lights coming from the projector. So, what causes us to fear that train coming at us, or lose our breath as if we really jump from that plane, or see and feel ourselves as if we are really in the middle of that epic battle, or love scene? What on that flat screen causes us to laugh, cry, scream, hate etc, as if it is real? The sequence of images imitating the motion triggers exactly the same sensations that would be triggered by the real events. The latter then "completes the picture" by triggering all the rest of the sensations corresponding to the real events that are in turn reflected in the brain's internal screen as a result. Thus, the cinema screen creates an illusion in our consciousness that we are witnessing or participating in real events because it creates exactly same sensations the real events would create. Consciousness is this never ending movie, where "I" is associated with the spectator. In addition, it is a multi-screen performance with hundreds of sub-screens that run various fluxes of consciousness, or channels, such as memories, history, news, daily events, business, science etc.

The precise definition of Consciousness is somewhat evading because there is no generally adopted understanding of what exactly it refers to. For example "being conscious" is equivalent to "being awake, or alert, or aware", and thus refers to the owner's condition of alert, as opposite, for example, to sleep. On the other hand, when one says that a thought is a product of consciousness there is no reference to the owner's condition at all. The thought could be as well produced in a sleep.

Nevertheless, it is still a sensation. It is similar to concepts that form the consciousness, and are based on sensations.

In addition, it depends on whether we want to include animal's consciousness in the definition. We already understood that human consciousness is just a complete set of concepts. However there are sensations that are "subconscious" and are not conceptualized. They correspond to the animal's part of our consciousness.

We therefore tend to define consciousness as a sensation of the brain's work, where the sensation is felt by the whole body. According to this definition it is neither illusion [3] nor mystery [4]. One can say that it is as much of an illusion as any image we see and as real as a pain that we feel. While a train running at us from the cinema screen is an illusion, the fear it causes is not.

While our brain is constantly at work, the feeling of this work is very limited and changing. It expands at certain times and shrinks at others. It is almost absent when we sleep. It can start thinking its own thoughts, and it does not know how the thinking is done.

Functionally, consciousness, as already mentioned, is a vision. Besides converting an "external" picture into an internal one, it "guesses" the details that are not in the external picture, it imagines, invents everything, and certainly lies. Surprisingly, our language, deceiving us in almost everything, sometimes yet adequately describes processes in the mind. Expressions such as "look at the root cause", "see the difference", "I see" (I understand), political "views", "foresight", "observation", etc use the terminology of optical vision although they refer to consciousness.

It is also understandable why it is so easy to deceive us. The desire to complete the picture, i.e. to have an explanation or theory for everything became our biological necessity. Our consciousness itself will gladly buy everything offered just to fill an empty space. We must know everything, understand everything, and have an explanation of everything. A false model, or concept, is better for us than their absence. This is what magicians and illusionists, swindlers and politicians, and other "kind" people actively use. Likewise, the simple concept is more readily acquired than the complicated one and more fiercely protected from extinction.

## On the origin of human intelligence.

Thus the human consciousness operates in terms of sensation-based concepts, which is an abstract thinking not found in animals. Therefore the important question is how this ability was acquired by our animal ancestors.

To answer this question let us recall how difficult it is for us to learn new things, to perceive new concepts, and especially to create new knowledge. In other words, the creation (or the acquisition) of a new concept is a difficult process requiring significant effort, exhausting repetitions, and hard mental work. Even the simplest concepts are not easy for a child to acquire. It is almost obvious that it is an extremely energy-consuming operation.

At the biological level, the creation of a new concept requires breaking the existing adjacency matrix of human connectome to insert a new "member" and establish new connections. This must be an extremely painful operation equivalent to surgery. Even though there are no pain receptors in the brain, it somehow must know about it and resist it with full force. Moreover, the more radical the new concept is, the more painful and energy-consuming is its creation.

This is the reason why many students prefer mindless memorization to conceptual learning, and why many people stop learning altogether at some point in their lives. Of course, there is a big difference between mastering everyday concepts and scientific ones. However, despite the huge initial energy consumption, mastering the apparatus of concepts is energetically beneficial over a long period of time, both for an individual and for the species as a whole, since it greatly facilitates the acquisition of knowledge in the future, facilitates the transfer of knowledge and allows predicting and forecasting.

My hypothesis is that human intelligence, i.e. module of concepts, has emerged in the chaotic mind of our animal ancestor as a result of a phase transition. The transition happens between the chaotic state of memorization and accumulation of facts (accumulative, animal phase) and the ordered state of generation of concepts (models, theories, systematic knowledge), with the ability to systematize, generalize, analyze, and predict (conceptual, human phase).

The accumulative phase can be also thought of as the experience phase, since accumulation of facts and accumulation of experience are the same. The conceptual phase is an intellectual phase. In other words, the work of human consciousness is a conversion of experience into conceptual knowledge. In physics we would call the first phase phenomenological or experimental phase, and the second one the theoretical phase. Each such conversion corresponds to a leap of intelligence. While accumulation of experience occurs gradually, the intelligence changes in steps (leaps).

The leap occurs at the moment when a new concept, a model, appears in consciousness. Then all the random "facts" fall into place, like spins in the Ising model. The newly created model immediately allows predicting, and depending on the result, it either confirms or disproves the model. This situation is familiar to every researcher, as well as to those students who were lucky enough to receive a systemic education, and acquire models in which the learned "facts" are fit. If the conceptual acquisition does not occur, then such an "education" is a sham. It is just a mindless accumulation of facts. Such "knowledge" in our time can be replaced by Google.

Similar to the individual transition, the acquisition of a concept, of a model, by a group of people is also a phase transition, though of a collective "mind". In order for the phase transition of one to become the phase transition of the entire community, it is necessary that the same phase transition occurs in the brains of all members. A collective phase transition requires considerable intellectual work by others, although to a lesser extent than by the "discoverer". The work of secondary members (colleagues who read the article of the "discoverer", then students, schoolchildren, if the result has become part of the curriculum) is facilitated by the fact that they know the result in advance. But nevertheless they must do some intellectual work in order to acquire this result, to master it in their mind. At the end the discovery becomes "their own". Sometimes this leads to (unintentional) scientific theft, when, as a result of such work, a person forgets that he received the idea from someone else.

Extrapolating into the remote past, it would be logical to assume that the emergence of the first concept, i.e. the very origin of the human intelligence was also a phase transition, just like all those we regularly make in our own brain (for those who does), or collectively, as a community.

I would venture to suggest that the biological reason for such events was the surplus of biomaterial, i.e. neurons that did not have enough roles for active participation in the process of external observation. As a result they turned to creating their own roles of "guessing", "completing" the details that are missing in the picture observed by their fellow partners, to looking for something that is not there, or even simply lying without malicious intent in order to "justify" their existence. In other words, too many actors for too little roles pushed the actors to create their own roles in the whole spectacle.

Besides the biological conditions for such a transition, there need to be a trigger for it, or, in the language of physics, an external field. The trigger can be any strong emotion or a random event. We can only guess what exactly the first concept was, and whether it resulted in the emergence of the very apparatus of concepts. I can suggest two scenarios.

**1. Evolutionary scenario: observation and guessing.**
As already mentioned, our brain, being a visual organ, is constantly working on generating pictures of what we see. It is easy to understand that the desire to recognize what is seen leads to guessing, coming up with details not actually seen in the real picture. You saw ears flickering in the grass and your imagination instantly draws one of the familiar concepts (tiger? wolf?), especially when it could be a danger. This is how concepts ("threat", "danger") and abstract thinking (inventing missing details) appear. The discovery of what you saw (or thought that you saw) has to be immediately reported to your herd in order to warn about the danger. The more details seen, the more reliable is the guess. Incidentally, humans are perhaps the only creatures that can instantly increase their viewing height by standing up from a sitting position and thereby see more details and verify the guess. Hence the need for upright walking may arise.

This would give a jump start to the appearance of the module of concepts, which could take millions of years to develop into a property that became stable and inheritable part of the nervous system.

I believe the latter is one of the most challenging questions for biology: How the ability to acquire concepts is encoded and inherited?

**2. Revolutionary scenario: fear and weapon.**
The previous hypothesis came straight from our newly gained understanding of the role of concepts, and what consciousness is. But we can also employ some facts we know from anthropology. For example, the fact that the first humans made weapons and tools, while animals do not make them. Therefore, it would be logical to assume that either weapons or tools were the first concepts. There are two considerations that could be helpful. The first is that one of our strongest emotions, if not the strongest, is fear. And the second is that inventing a new concept is not just difficult as mentioned above, but it is also scary. Fear sometimes makes us think and look for an escape from a dangerous situation. It also helps to overcome other less significant fears. For example fear for life may help overcome fear of pain. And what can one think of under the fear, especially in a situation of fear for life? The answer is weapon!

This is not just about the use of weapon, but about the emergence of the very concept of weapon. Monkeys, according to various sources, can also use a stick as a weapon, and not just as a tool for knocking down bananas. In addition, it is alleged that monkeys also use stone throwing as a weapon. But unlike ancient human, they do not make a spear from a stick, which they then carry with them almost all the time, to equalize their chances in confrontation with a lion.

For this to happen, something extraordinary had to take place that caused our ancestor to invent weapons and thereby make a phase transition in his brain, thus becoming first human. For example, the fear of imminent death in the paws and teeth of a powerful predator suddenly ended by a well-used stick with a sharp end that miraculously killed the attacker. The tremendous change of emotion from the mortal fear to the joy of salvation is quite capable of making the monkey realize (and remember) that such a stick can save him next time too. Therefore, the stick must be carefully selected, sharpened and always carried, which all together constitute the concept of weapon.

What is more important, this is accompanied by the emergence of the first logical operator (the identity operator), the operator of identifying a new concept, in this case 'weapon' as a means of the defense (and, of course, of the assault):

$$\text{Stick} = \text{weapon} \qquad (1)$$

This gives an incentive to look for other useful concepts, by means of identifying various objects as 'weapons' or 'tools'.

Yet another important consequence of (1) is the causal connection between the demise of the attacker and the lucky use of weapon. For humans this connection is obvious, but for an animal it is a huge leap.

I dare to claim that a discovery like (1) is more revolutionary than Einstein's $E = mc^2$. Not only because of its immediate practical value but first of all because it made a revolutionary change in the brain of our ancestor, and in fact converted him into a modern thinking human. It became the beginning of creation of the module of concepts, and prompted the next generations to look for connections between seemingly unrelated things.

From the point of view of the survival of the species, this is followed by 1) the need to free hands, and thus become bipedal, 2) the ability to move out of the jungle to the plain, where bipedal walking allows one to see further, and where the tail becomes an atavism, 3) the need to think about improving weapons and inventing new ones, thus turning the brain into the weapon.

In this scenario, the fear and the invention of weapons turned our animal predecessor into a human, and radically changed his appearance and status in the animal world. It made him a bipedal, large-headed, tailless, armed creature, the strongest in the animal kingdom, and lifted him to the top of the food chain. The intelligence defeated strength and even experience.

Once again, these are just two among many possible hypotheses.

As we very well know the ability to conceptualize, i.e. the module of concepts is inherited by human children. They do not have the actual concepts at birth but they do have an ability to acquire them, unlike animals that cannot be taught any concept at all. I therefore tend to think that scenario 1 was prevailing for millions of years, preceding scenario 2 that either occurred or not only long time after that.

There is no doubt that the emergence of the module of concepts created a significant need for language, since every new concept requires notation, tabulation, indexing. With the growing number and complexity of concepts, more and more complex sounds are required to tabulate them. This may explain how language was born. It allows transferring knowledge, training offspring, uniting larger groups, and therefore organizing collective protection, hunting, and raising children. It frees females, elders, and tool makers. Communities are emerging. For a further history of Sapiens, one can read Harari [1].

The emergence of spoken language finally consolidated this transition and allowed a human to become a social creature, customarily called Homo Sapiens while the correct name would be Homo Dolosus (lying, deceiving man).

From this moment on, evolutionary development, which works on the scale of millions of years, can be considered complete. A chain of revolutions has begun. On a historical scale, two million years passed from the emergence of the first logical operator in the pre-human brain that identified a stick as a spear, to the identification of energy as a mass and to the creation of an atomic bomb is an instant.

Undoubtedly, some animals behave as if they also have certain intelligence. They have models of behavior developed over millions years. But it is also obvious that these models appeared not as a result of intelligence, but as a result of accumulated experience and natural selection. The species whose behavioral models do not meet the tasks of the survival in given conditions die out.

Recently, I learned from Sasha Gorsky that a study conducted by him and co-authors [5] indicates that the human brain is constantly in a near-critical state, i.e. near the phase transition. This indirectly confirms the hypothesis that the emergence of a concept (and this is what our consciousness is constantly busy with, or at least is going to do) is a phase transition.

## Conclusions:

1. The fundamental elementary constituents of human consciousness are concepts. Biologically, a concept is a sensation developed and learned by an individual's nervous system.

2. Consciousness, physically and biologically, is a sensation of the brain's work. Functionally, consciousness is a vision.
3. The emergence of a new concept is a phase transition.
4. Concepts are the only principal difference between humans and animals. Emergence of the module of concepts meant emergence of humans.
5. The language emerged as a necessity to denote a growing number of new concepts, as a tool of transferring them to the offspring, and as a means of communication within community.
6. Memory is the association of sensations with corresponding image, a thought, or a concept.

**Acknowledgements.**
I am grateful to late Professor N.N. Meiman with whom I shared an office in 1991-1993 at Tel Aviv University and had numerous conversations about many aspects of the scientific mind, and the peculiarity of scientific ways. His observations came from years of collaboration and friendship with many famous physicists and gave me invaluable insights for this writing. I am grateful to Lev Neyman for discussions during this research. I am grateful to Professor Vitaly Polunovsky for enlightening me on certain aspects of modern genetics and for discussions of various issues related to this writing.